\documentclass[aps,prl,twocolumn,letterpaper,superscriptaddress]{revtex4-2}
\usepackage{graphicx}
\usepackage{amsmath}
\usepackage{amssymb}
\usepackage{url}
\usepackage{hyperref}

\begin{document}
\title{Wegner model on a tree graph: $\mathrm{U}(1)$ symmetry breaking and a\\ non-standard phase of disordered electronic matter}
\author{Julian Arenz}
\affiliation{Institut f\"ur Theoretische Physik, Universit\"at zu K\"oln, Z\"ulpicher Str. 77a, 50937 K\"oln, Germany}
\author{Martin R. Zirnbauer}
\affiliation{Institut f\"ur Theoretische Physik, Universit\"at zu K\"oln, Z\"ulpicher Str. 77a, 50937 K\"oln, Germany}
\date{\today}

\begin{abstract}
Assuming the self-consistent theory of localization due to Abou-Chacra {\it et al.}, we solve the $N=1$ Wegner model in the regime of strong disorder and high dimension. In the process, we uncover a non-standard electronic phase with spontaneously broken $\mathrm{U}(1)$ symmetry --- the missing field-theory basis underlying phenomena associated with fractal eigenstates and singular continuous spectra.
\end{abstract}

\maketitle

\section{Introduction}
Disordered electrons undergo a quantum phase transition, known as the Anderson transition, from a metallic state at weak disorder to an insulating state at strong disorder. In the present Letter, we address the situation of electrons treated in the non-interacting (or mean-field) approximation. Following an influential paper \cite{GangOfFour}, it has become standard to assume for the Anderson transition a one-parameter scaling hypothesis. Early support for that hypothesis came from the renormalization group (RG) treatment of an effective field theory, the nonlinear $\sigma$ model \cite{Weg79a}, analyzed by $(2+\epsilon)$-expansion in the weak-coupling regime near two space dimensions. It is widely believed at present that the nonlinear $\sigma$ model continues to apply in high dimensions. There is the caveat, however, that the Anderson transition moves into the strong disorder (or strong-coupling) regime with increasing dimension, and there exists no proof of renormalizability of the nonlinear $\sigma$ model at strong coupling. In fact, we are going to argue here that the standard scenario for Anderson transitions in high dimension needs revision.

The standard scenario posits that there exist two (and no more than two) phases: (i) the metallic state with absolutely continuous energy spectrum and spatially extended eigenfunctions, and (ii) the insulating state with pure point spectrum and localized eigenfunctions. Now some time ago, it was proposed \cite{AGKL1997, KAI2018} that there may exist a third phase, called NEE for non-ergodic extended, where the electron eigenstates are neither metallic nor insulating, but fractal. This proposal is borne out for simple models of random-matrix type \cite{KKCA2015} but has been refuted for the Anderson model on tree-like graphs \cite{TMS2016, BHT2022, Lemarie2022, SLS2022}.

How can we put the possibility of a third phase on solid theoretical ground? We can discern two avenues for that. For one, it is known that a random Schr\"odinger operator (or any self-adjoint Hamiltonian for disordered electrons) in the infinite-volume limit may support three (not just two) types of spectrum: apart from absolutely continuous (ac) and pure point (pp) spectrum, there is also the possibility of singular continuous (sc) spectrum \cite{ReedSimon, DJMS94}. Now, a rather natural idea is to link the characteristic features of fractal eigenstates with those of sc spectrum and, indeed, this link has been pursued in a recent preprint \cite{AK2023}. Another avenue is to establish the third phase as a true thermodynamic phase in the sense of Landau, by pinpointing a field-theory scenario in which a global symmetry is spontaneously broken by the formation of an order parameter. Initiating the framework for that potent scenario will be the thrust of our Letter.

The plan here is to investigate a variant of the standard Anderson model, namely the Wegner $N$-orbital model \cite{Weg79b} of type $A$ of the Tenfold Way \cite{HHZ}. While previous treatments were mostly concerned with the metallic limit for large $N$, our focus here is on the model for $N = 1$ (i.e.\ with a single orbital per site) in the regime of strong disorder and high mobility. We shall solve that model in the approximation afforded by the self-consistency equation of Abou-Chacra, Anderson, and Thouless (AAT) \cite{AAT1973}. Exact on the Bethe lattice (an infinite regular tree), the AAT approximation is expected to capture the relevant physics in high dimension. Our treatment based on AAT is quite elementary, as the further simplifications for the $N=1$ Wegner model allow us to avoid the supersymmetry formalism usually employed in the present context.

As for the outline, we first present our main message as a conjectured RG flow diagram for Anderson transitions at strong coupling.
We then undertake to study the Fourier-Laplace transform, $Y$, of the probability distribution for the reciprocal local Green's function (retarded and advanced) in infinite volume. Adopting the AAT approximation, we set up a self-consistency equation for $Y$. That equation is invariant under the action of a Lorentzian group $G^\prime$. From past work we know that the solution $Y_{\rm pp}$ in the insulator phase is $G^\prime$-invariant, while the metallic solution $Y_{\rm ac}$ breaks the $G^\prime$-symmetry maximally. Our main result is a third type of solution, $Y_{\rm sc}$, which breaks the $G^\prime$-symmetry \emph{partially}. Expected to be unstable with respect to generic perturbations in low dimension, that solution  $Y_{\rm sc}$ is stable for a finite range of disorder strengths in high dimension. By virtue of the stability of $Y_{\rm sc}$, there must exist a third phase, distinct in Landau's sense from the known ones (pp, ac) by a distinct form of symmetry breaking. In the conclusions of the paper, we propose that our novel symmetry-breaking scenario provides the missing field-theoretical foundation on which to harvest the phenomena associated with fractal energy eigenstates and singular continuous spectra.

\section{Conjectured RG Flow Diagram}
Numerical studies \cite{BHT2022, Lemarie2022, SLS2022} agree that the Anderson model on a variety of tree-like graphs exhibits a single Anderson transition, with no third phase intervening. There is also agreement that the numerical findings defy the traditional one-parameter scaling hypothesis, as two (not one) relevant parameters appear. However, there fails to be agreement as to the details of the critical behavior and the specific two-parameter scaling theory to adopt. Ref.\ \cite{Lemarie2022} focuses on the insulator side of the transition and argues for behavior of Kosterlitz-Thouless (KT) type. In contrast, Ref.\ \cite{SLS2022} focuses on the metal side and finds two length scales with power-law criticality (not KT).

\begin{figure}
    \centering
    \includegraphics[width=8cm]{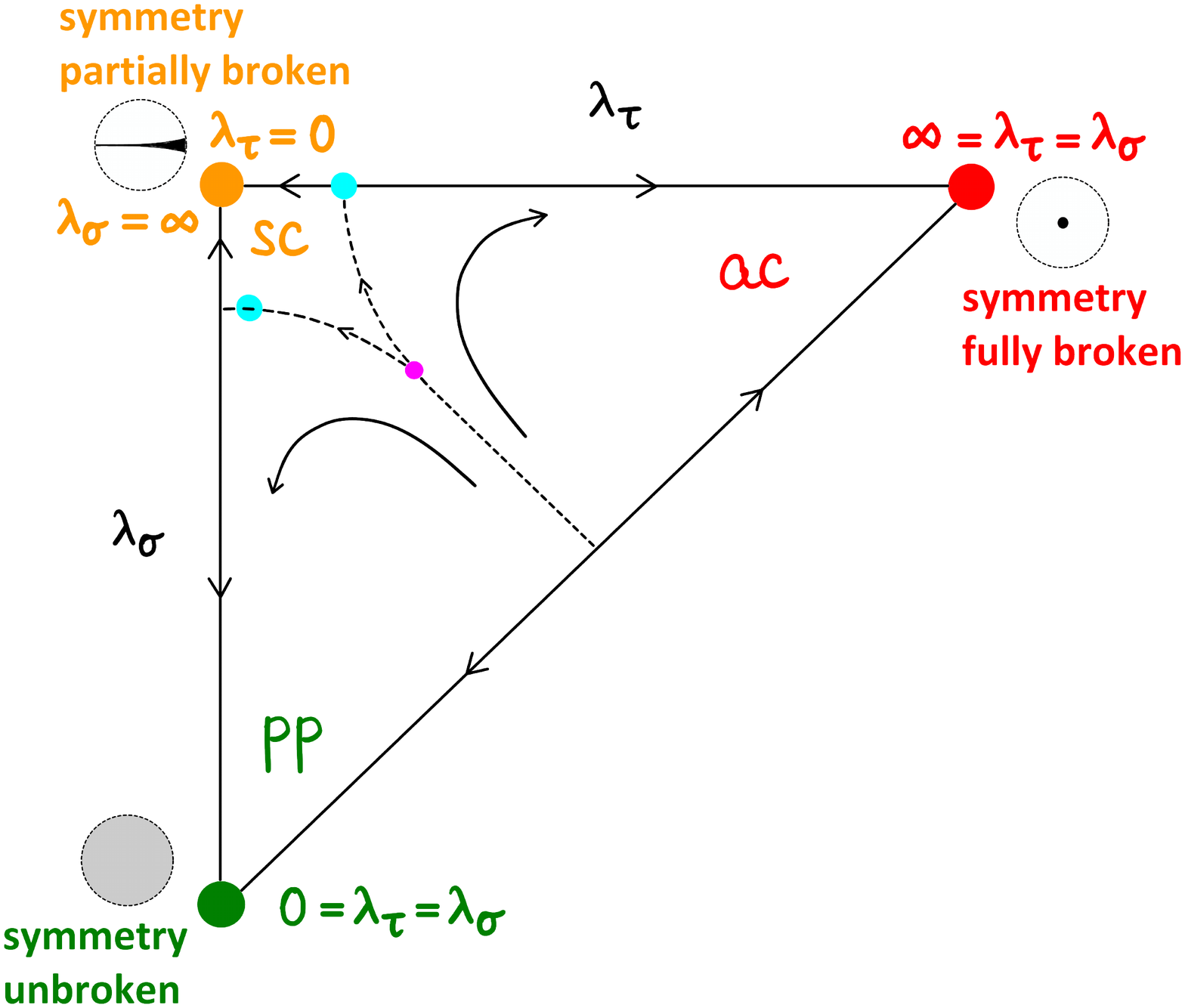}
    \caption{Schematic renormalization group (RG) flow diagram for a two-parameter field theory underlying Anderson transitions at strong coupling (conjecture). The field theory has a global Lorentzian symmetry. The reciprocal coupling $\lambda_\tau$ (resp.\ $\lambda_\sigma$) is the stiffness for field fluctuations tangent to Lorentz boost directions (resp.\ tangent to motions of compact type) of the target space. In addition to the two known RG-fixed points ``ac'' ($\lambda_\sigma = \lambda_\tau = \infty:$ metal) and ``pp" ($\lambda_\sigma = \lambda_\tau = 0:$ insulator), there is a third `trivial' RG-fixed point,``s'', at $\lambda_\sigma = \infty$, $\lambda_\tau = 0$. In a certain parameter range, the $N=1$ Wegner model is argued to flow to that fixed point.} \label{fig:F1}
\end{figure}

We take this persistent disagreement as a strong indication that a simple scaling theory covering both sides of the transition does not exist. To resolve the conundrum, we propose that the multi-critical point of a renormalizable field theory with reciprocal couplings $\lambda_\sigma > \lambda_\tau$ (superseding the nonlinear $\sigma$ model, where $\lambda_\sigma = \lambda_\tau$) splits the critical renormalization group (RG) flow, thereby channeling the two sides of the Anderson transition to \emph{separate} critical points leading to different critical behaviors (Fig.\ \ref{fig:F1}). Shadowed by the multi-critical point, there is a trivial RG-fixed point ($\lambda_\tau = 0$, $\lambda_\sigma = \infty$), which serves as the point of attraction for a phase (sc) distinct from the established phases (pp and ac). In the sequel, we shall substantiate this conjecture by solving the $N=1$ Wegner model in the AAT self-consistent approximation.

\section{Model and Technique Used}
The object studied in this paper is a random Hamiltonian, $H$, of Schr\"odinger type in the discrete setting of a graph, $\mathbb{G}$. Denoting the sites of $\mathbb{G}$ by $n$ we express $H$ as
\begin{equation}\label{eq:Wegner1}
    H = \sum h_{n,n^\prime} \, c_{n^{\vphantom{\prime}}}^\dagger c_{n^\prime}^{\vphantom{\dagger}} \,.
\end{equation}
Although $H$ is presented here in basis-independent form using particle creation and annihilation operators, we will analyze it as an operator ($h$) on the single-particle Hilbert space $\ell^2(\mathbb{G})$. All Hamiltonian matrix elements are taken to be complex Gaussian independent random variables (subject to the Hermiticity condition $h_{n^\prime,n} = \overline{h_{n,n^\prime}}$) with zero mean and variance
\begin{equation}\label{eq:problaW}
    \mathbb{E} \left( h_{n,n^\prime} h_{l,l^\prime} \right) = \delta_{n l^\prime} \delta_{n^\prime l} \left( \delta_{n n^\prime} w_V^2 + A_{n n^\prime} w_T^2 \right) ,
\end{equation}
where $A_{n n^\prime} = A_{n^\prime n} \in \{ 0 , 1 \}$ is the adjacency matrix of the graph $\mathbb{G}$. The parameter $w_V^2$ sets the scale of variation of the on-site potentials $h_{n,n} = V_n$, while $w_T^2$ sets the (kinetic-energy) scale for hopping between nearest-neighbor pairs. The model so defined will be referred to simply as the ``Wegner model'' (it is officially known as the Wegner $N$-orbital model of symmetry class $A$ with $N = 1$, i.e.\ a single orbital per site). Our interest here is in the regime of strong diagonal disorder, $w_V^2 \gg w_T^2$, on a graph $\mathbb{G}$ with high coordination number, $K+1$.

By the probability law (\ref{eq:problaW}), the ensemble of Hamiltonians (\ref{eq:Wegner1}) is invariant under local $\mathrm{U}(1)$ phase rotations
\begin{equation}
    c_n^\dagger \mapsto   c_n^\dagger \mathrm{e}^{\mathrm{i} \alpha_n} , \quad
    c_{n^\prime} \mapsto \mathrm{e}^{- \mathrm{i} \alpha_{n^\prime}}  c_{n^\prime}  .
\end{equation}
This property is not shared by the Anderson model (where the hopping matrix elements are non-random and constant) but is expected to emerge \cite{PS82} under renormalization on large length scales. Its attractive feature is that it facilitates the introduction of matrix variables.

\subsection{AAT self-consistency equation}
The analysis starts from the retarded Green's function, $G \equiv G^+$, which, for any energy $E$, is defined by
\begin{equation}
    \sum_n (E + \mathrm{i} \varepsilon - h)_{n_1 , n} \, {G}_{n , n_2} = \delta_{n_1 n_2}  \quad (\varepsilon > 0) .
\end{equation}
By simple linear algebra, the reciprocal of the diagonal matrix element ${G}_{n,n} \equiv {G}_n$ can be expressed as
\begin{equation}
    \frac{1}{{G}_n} = E + \mathrm{i}\varepsilon - h_{n,n} - \sum_{n_1 , n_2} h_{n, n_1} {G}_{n_1 , n_2}^\prime h_{n_2 , n} \,,
\end{equation}
where ${G}^\prime$ is the (retarded) Green's function of the system with the site $n$ omitted.

We now adopt an approximation scheme introduced by Abou-Chacra, Anderson, and Thouless \cite{AAT1973}. In that scheme (AAT), one omits all off-diagonal terms (and also one of the diagonal terms) from the sum on the right-hand side, and one assumes that the law of the random variables ${G}_{n_1 = n_2}^\prime$ in the resulting expression can be approximated by that of ${G}_n$. The result of these approximations is the AAT self-consistency equation:
\begin{equation}\label{eq:AAT}
    \frac{1}{{G}_n}\stackrel{!}{=} E + \mathrm{i}\varepsilon - h_{n,n} - \sum_{n^\prime (\not= n)} |h_{n, n^\prime}|^2 {G}_{n^\prime}.
\end{equation}
Note that the sum on the r.h.s.\ runs over $K$ terms (for graph coordination number $K+1$). The notation $\stackrel{!}{=}$ is to say that both sides, as random variables, obey the same law in probability, which means the following. (i) There exists a probability distribution for the complex random variable $1/ G_n$; (ii) if we draw $K$ independent values for the $1/{G}_{n^\prime}$ from that same distribution and (iii) we independently draw values for $h_{n,n}$ and $h_{n,n^\prime}$ from their zero-centered Gaussian distributions with variances $w_V^2$ resp.\ $w_T^2$, then (iv) by forming the expression on the r.h.s.\ of Eq.\ (\ref{eq:AAT}) we obtain a value that is statistically the same as a value drawn from the distribution of $1/G_n$.

The AAT approximation is exact on a Bethe lattice and is expected to be reasonable in high dimension.

\subsection{Fourier-Laplace transform}
Our goal now is to use Eq.\ (\ref{eq:AAT}) in order to find the probability distribution of $G_n$ (for a fixed energy $E$ and coordination number $K+1$) from the given distributions of $h_{n,n}$ and $h_{n,n^\prime}$. Of course, with enough computer power this goal could be achieved directly by numerical simulation. Our intention, however, is to develop an analytical understanding and, in particular, pin down the law for the imaginary part of $G_n$ (and a matrix generalization thereof). That law exhibits critical behavior near the Anderson transition, going from $\delta$-type singular for localized states to smooth for extended states. As is often the case, that singular distribution is easier to handle analytically via its characteristic function or Fourier-Laplace transform. With this motivation, we proceed as follows.

To begin, observing that $\mathrm{Im} (1/{G}_n) > 0$ (for $\varepsilon > 0$), we introduce the transform
\begin{equation}
    Z(p) := \mathbb{E} \big( \mathrm{e}^{\mathrm{i} p {G}_n^{-1}} \big) \quad (p > 0) ,
\end{equation}
where $\mathbb{E}(\ldots)$ means the expectation value with respect to the distribution of the random variable $1/{G}_n$. Assuming the AAT approximation (\ref{eq:AAT}), one derives \cite{AZ23} for the function $Z(p)$ a nonlinear integral equation:
 \begin{align}\label{eq:eqn4Z}
    Z(p) &= \mathrm{e}^{\mathrm{i}p(E + \mathrm{i}\varepsilon) - w_V^2 p^2 / 2 } \cr &\times \left( - \int\nolimits_{0}^{\infty} dp^\prime \, \mathrm{e}^{-w_T^2 p p^\prime} \frac{\partial}{\partial p^\prime} Z(p^\prime) \right)^{K} .
\end{align}
This equation is relatively easy to analyze. Its solution is a smooth function interpolating between $Z(p = 0) = \mathbb{E}(1) = 1$ and $Z(p \to \infty) = 0$. For energy $E = 0$ the function $Z(p)$ is real-valued, positive, and monotonic. $Z(p)$ extends to negative values of $p$ by switching the signs of $p^\prime$ and $\varepsilon$ (giving the advanced Green's function).

\subsection{Local Green's function generalized}
The function $Z(p)$, while basic to the analysis, has the shortcoming that it does not isolate the imaginary part of $G_n^{-1}$ from the real part; for that reason, it does not exhibit Anderson-transition critical behavior. To see any criticality, the advanced Green's function ($\varepsilon < 0$) must be brought into play along with the retarded one.

Let us consider the matrix Green's function ($\varepsilon > 0$)
\begin{equation}
    \mathcal{G} = \begin{pmatrix} E + \mathrm{i}\varepsilon - h & \mathrm{i}\beta \cr \mathrm{i}\beta &E - \mathrm{i}\varepsilon - h \end{pmatrix}^{-1} ,
\end{equation}
where for an enhanced perspective we have added a site-diagonal perturbation $\beta_{n,n^\prime} = \delta_{n,n^\prime} b_n$ by Gaussian i.i.d.\ real random variables $b_n$ with zero mean and variance $w_\beta^2$. The latter will be taken to zero in our demonstration of a novel form of spontaneous symmetry breaking.

We again focus on the local Green's function, $\mathcal{G}_n \equiv \mathcal{G}_{n,n}$, which is now a $2 \times 2$ matrix. The inverse matrix, $\mathcal{G}_n^{-1}$, has 3 real degrees of freedom $A_n$, $B_n$, and $D_n$:
\begin{equation}\label{eq:strucG}
    \mathcal{G}_n^{-1} = \begin{pmatrix} A_n + \mathrm{i} D_n & \mathrm{i} B_n \cr \mathrm{i} B_n &A_n - \mathrm{i} D_n \end{pmatrix} ,
\end{equation}
with $D_n > 0$. Note that $\mathrm{Det}(\mathcal{G}_n^{-1}) = A_n^2 + D_n^2 + B_n^2 > 0$.

The requisite generalization of the function $Z(p)$ is a Fourier-Laplace transform for the distribution of $\mathcal{G}_n^{-1}$:
\begin{equation}
    Y(Q_n) := \mathbb{E} \big( \mathrm{e}^{\mathrm{i} \mathrm{Tr}\, Q_n \mathcal{G}_n^{-1}} \big) ,
\end{equation}
where the conjugate variable is now a $2 \times 2$ matrix
\begin{equation}\label{eq:rankone}
    Q_n = \begin{pmatrix} u_n \bar{u}_n &- u_n \bar{v}_n \cr v_n \bar{u}_n &- v_n \bar{v}_n \end{pmatrix} \quad (u_n , v_n \in \mathbb{C})
\end{equation}
with the properties $Q_n = \sigma_3 Q_n^\dagger \sigma_3$ and $\mathrm{Det}(Q_n) = 0$.

$Y$ contains substantial information about the random Hamiltonian $H$ and the observables associated to it. Indeed, by integrating $Y$ against polynomials in $u_n$ and $v_n$ one obtains disorder averages of products of retarded and advanced local Green's functions. {}From the latter one can extract spectral and wavefunction observables.

We note that the matrix Green's function $\mathcal{G}^{-1}$ commutes with $\sigma_3$ for $w_\beta^2 \to 0$. In that limit, one might consider it reasonable to assume for $Y$ a $\mathrm{U}(1)$ symmetry
\begin{equation}\label{eq:U(1)}
    Y(Q_n) \stackrel{?}{=} Y(\mathrm{e}^{\mathrm{i}\alpha \sigma_3} Q_n \, \mathrm{e}^{-\mathrm{i}\alpha \sigma_3}) .
\end{equation}
Remarkably, we find that the symmetry (\ref{eq:U(1)}) may be broken spontaneously, for strong disorder in high dimension. The existence of such a possibility was overlooked in \cite{MF1992}.

\subsection{AAT for the matrix Green's function}
AAT self-consistency (transcribed to the matrix-valued function $\mathcal{G}$) still leads to an equation for $Y(Q_n \equiv Q)$:
\begin{align}\label{eq:eqn4Y}
    Y(Q) &= \mathrm{e}^{\mathrm{i}\, \mathrm{Tr}\, Q (E + \mathrm{i}\varepsilon \sigma_3) - \mathrm{Tr} \, (w_V^2 Q^2 - w_\beta^2 \sigma_1 Q \sigma_1 Q)/2} \cr &\times \left( \int dQ^\prime \, \mathrm{e}^{- w_T^2 \mathrm{Tr}\, Q Q^\prime} \Delta Y(Q^\prime) \right)^{K} , \\
    &dQ^\prime = \pi^{-2} d^2u^\prime \, d^2v^\prime \quad (u^\prime, v^\prime \in \mathbb{C}) , \nonumber
\end{align}
where $\Delta$, a differential operator of second order in the matrix elements of $Q^\prime$, is defined by
\begin{equation}
    \mathrm{Det} (\mathcal{G}^{-1}) \, \mathrm{e}^{\mathrm{i} \mathrm{Tr}\, Q \mathcal{G}^{-1}} =
    \Delta\, \mathrm{e}^{\mathrm{i} \mathrm{Tr}\, Q \mathcal{G}^{-1}} .
\end{equation}
The proof of Eq.\ (\ref{eq:eqn4Y}) is given in \cite{AZ23}. We remark that one usually represents  $\mathrm{Det} (\mathcal{G}^{-1})$ as a Gaussian integral over Grassmann variables, thereby entering the world of super-calculus. Such technicalities can be avoided here to some extent, by virtue of the simplifications due to the Wegner model treated in the AAT approximation.

To be precise, we must caution that Eq.\ (\ref{eq:eqn4Y}) is not closed in general, as the definition of $\Delta$ requires a matrix $Q$ of full rank while $Q_n$ in (\ref{eq:rankone}) is rank-deficient. Fortunately, this issue can be fixed in the cases treated below.

Next we observe that the matrix $Q_n \equiv Q$ in Eq.\ (\ref{eq:rankone}) does not depend on the overall phase of $u$ and $v$. By that token, the integral (\ref{eq:eqn4Y}) can be reduced to 3 degrees of freedom. For some purposes, a good parametrization is
\begin{equation}\label{eq:Q-cart}
    Q = x_0 \cdot 1 + x_3 \sigma_3 + \mathrm{i}(x_1 \sigma_1 + x_2 \sigma_2)
\end{equation}
by 3 real variables $x_0, x_1, x_2$ and $x_3 = + \sqrt{x_0^2 + x_1^2 + x_2^2}$ (which, incidentally, is the on-shell condition in $2+1$ dimensions for a relativistic particle with invariant mass $|x_0|$, momentum components $(x_1, x_2)$ and energy $x_3 \geq 0$). The integration measure $dQ = \pi^{-2} d^2u \, d^2v$ reduces to
\begin{equation}
    d\mu(Q) = \pi^{-1} dx_0 \frac{dx_1 dx_2}{x_3} \,.
\end{equation}

The function $Y(Q)$ extends the function $Z(p)$ in the following sense. Let $w_\beta^2 = 0$. When we set $v \equiv 0$ in $Q$, the kernel $\mathrm{e}^{-w_T^2 \mathrm{Tr}\, Q Q^\prime}$ reduces to $\mathrm{e}^{-w_T^2 p p^\prime}$ ($p \equiv |u|^2$), and the equation for $Y$ collapses to that for $Z$. (The same happens on setting $u \equiv 0$.) By consequence, some boundary values of $Y$ are prescribed by $Z$:
\begin{equation}
    Y \vert_{Q = \begin{pmatrix} p &0\cr 0 &0 \end{pmatrix}} = Z(p) , \quad
    Y \vert_{Q = \begin{pmatrix} 0 &0\cr 0 &-p \end{pmatrix}} = Z(-p) .
\end{equation}

\section{Analysis: symmetry breaking}
We open with a remark on symmetry: in the limit of $\varepsilon, w_\beta^2 \to 0$, the self-consistency equation (\ref{eq:eqn4Y}) becomes invariant under global pseudo-unitary transformations
\begin{align}\label{eq:pseudo}
    (Q,Q^\prime) &\mapsto (g Q g^{-1}, g Q^\prime g^{-1}) , \cr g &= \sigma_3 (g^{-1})^\dagger \sigma_3 \in \mathrm{SU}(1,1) .
\end{align}
The unusual features of the Anderson transition can be attributed to the group $G \equiv \mathrm{SU}(1,1)$ being non-compact. We note that the $G$-action by conjugation in (\ref{eq:pseudo}) is effectively by the Lorentz group $\mathrm{SO}(1,2) \equiv G^\prime \cong G / \mathbb{Z}_2\,$.

While Eq.\ (\ref{eq:eqn4Y}) is manifestly $G$-invariant, the solution $Y$ may or may not have that symmetry. Different phases of disordered electronic matter correspond to different types of spontaneous $G$-symmetry breaking. We proceed to exhibit three different types of solution.

\subsection{Solution of pp-type}
For $\varepsilon, w_\beta^2 \to 0$, Eq.\ (\ref{eq:eqn4Y}) for $Y$ has a trivial solution:
\begin{equation}
    Y_{\rm pp} (Q) = Z(\mathrm{Tr}\, Q) = Y_{\rm pp}(g Q g^{-1}) ,
\end{equation}
which is symmetry-unbroken, i.e.\ constant along the orbits of the symmetry group $\mathrm{SU}(1,1)$. Given that $Y_{\rm pp}$ depends only on $\mathrm{Tr}\, Q = 2 x_0$, we may close Eq.\ (\ref{eq:eqn4Y}) by taking $4 \Delta = - \partial^2 / \partial x_0^2$. The solution $Y_{\rm pp}$ is stable with respect to generic perturbations for strong diagonal disorder ($w_V^2 \gg w_T^2$) and $K$ not too large. Using this solution to compute the expectation $\mathbb{E}(|G_n|^2)$ of the squared local Green's function, one finds the result to be divergent (by the infinite volume of the symmetry orbits). The divergence signals that $G_n(E)$ has poles on the real energy axis. In this case, the spectral measure is of pure-point type, corresponding to localized eigenfunctions for $H$. 

\subsection{Solution of ac-type}
In the opposite limit of weak disorder ($w_V^2 \ll w_T^2$) and $K$ large, the solution $Y_{\rm pp}$ is unstable under iterations of Eq.\ (\ref{eq:eqn4Y}). The stable solution, $Y_{\rm ac}\,$, is symmetry-breaking. Assuming that $Y$ is independent of $\mathrm{arg}(\bar{u} v)$, we integrate over that phase. Then the kernel $\mathrm{e}^{- w_T^2 \mathrm{Tr}\, Q Q^\prime}$ reduces to
\begin{equation}
    \mathrm{e}^{-w_T^2 (p p^\prime + q q^\prime)} \, I_0(2w_T^2 \sqrt{p q p^\prime q^\prime})
\end{equation}
(modified Bessel function $I_0$) for $p = x_3 + x_0$, $q = x_3 - x_0$, and the self-consistency equation (\ref{eq:eqn4Y}) is closed by
\begin{equation}
    \Delta = \frac{1}{4} \left( \frac{\partial^2}{\partial x_3^2} - \frac{\partial^2}{\partial x_0^2} \right) = \frac{\partial^2}{\partial p \partial q}\,.
\end{equation}
Long known from the work of \cite{MF1992, AbelKlein98, AW2013}, the solution $Y_{\rm ac}$ exhibits rapid decay in the variables $p$ and $q$, thereby breaking the $\mathrm{SO}(1,2)$ Lorentz boost symmetries generated by $\sigma_1$ and $\sigma_2$. By virtue of that decay, the expectation $\mathbb{E}(|G_n|^{r})$ is finite for any $r \in 2\mathbb{N}$ (actually any real $r$), which is the hallmark of absolutely continuous spectral measure and ergodic eigenfunctions of the Hamiltonian.

\subsection{Solution of sc-type}
Focusing on the center $E = 0$ of the energy band for simplicity, we finally describe a solution $Y_{\rm sc}$ of novel type, which stably exists for strong disorder and large $K$.

There is a clash between two opposing tendencies in Eq.\ (\ref{eq:eqn4Y}): integration against the kernel $\mathrm{e}^{-w_T^2 \mathrm{Tr}\, Q Q^\prime}$ \emph{broadens} the function $Y$ in all $G$-symmetry directions, whereas the ensuing step of raising $Y$ to the power of $K$ \emph{narrows} it (note $0 \leq Y \leq 1$). The former operation tends to restore the Lorentz group symmetry $\mathrm{SO}(1,2)$, while the latter tends to break it spontaneously. Either tendency may prevail (cf.\ the pp- and ac-type scenarios above).

Now in the strong-coupling limit ($w_V^2 \gg w_T^2$), fluctuations of the ``mass'' variable $x_0$ are strongly suppressed by the factor $\mathrm{e}^{- w_V^2 {\rm Tr}\, Q^2 /2}$ in (\ref{eq:eqn4Y}). Setting $x_0 = 0$ in leading approximation, we get an integral kernel of the form
\begin{equation}
    \exp\left( - 2 w_T^2 (x_3^{\vphantom{\prime}} x_3^\prime - x_1^{\vphantom{\prime}} x_1^\prime - x_2^{\vphantom{\prime}} x_2^\prime) \right),
\end{equation}
whose range is \emph{infinite} (!) in the $\mathrm{SO}(1,2)$ Lorentz boost directions (determined by $x_3^{\vphantom{\prime}} x_3^\prime - x_1^{\vphantom{\prime}} x_1^\prime - x_2^{\vphantom{\prime}} x_2^\prime = 0$) but \emph{finite} in the direction of $\mathrm{SO}(2)$ rotations in the $x_1 x_2$ plane. When combined with the narrowing effect from the operation $f \mapsto f^{K}$, this strong anisotropy opens the striking possibility for a Lorentz boost symmetry to be unbroken but rotational symmetry to be broken.

In order to demonstrate that this striking scenario is borne out, we adopt a horospherical coordinate system
\begin{align}
    x_0 = \mathrm{Im}(a), \;\; x_1 = \mathrm{Re}(a) , \;\; x_2 = \big( |a|^2 \mathrm{e}^{- \varphi} - \mathrm{e}^\varphi \big) / 2 ,
\end{align}
given by variables $a \in \mathbb{C}$ and $\varphi \in \mathbb{R}$. Note that $x_3 = ( |a|^2 \mathrm{e}^{- \varphi} + \mathrm{e}^\varphi)/2$ with $\varphi$ the parameter of a Lorentz boost in the $x_2 x_3$ plane. The integration measure turns into $d\mu(Q) = \pi^{-1} d^2 a \, d\varphi$, and the kernel becomes
\begin{equation}
    \mathrm{e}^{- w_T^2 {\rm Tr}\, Q Q^\prime} = \mathrm{e}^{w_T^2 \big( 2 \mathrm{Re}(aa^\prime) - |a|^2 \mathrm{e}^{\varphi^\prime - \varphi}  - |a^\prime|^2 \mathrm{e}^{\varphi - \varphi^\prime} \big) } .
\end{equation}

We now show that the AAT equation (\ref{eq:eqn4Y}) has solutions $Y_{\rm sc}(a)$ independent of the Lorentz boost parameter $\varphi$. For such solutions, the equation reduces to
\begin{align}\label{eq:eqn4Ysc}
    Y_{\rm sc} (a) &= \mathrm{e}^{- w^2 \mathrm{Im}^2 (a) - (w_\beta^2 / w_T^2)\, \mathrm{Re}^2 (a)} \cr
    &\times \left( \int \frac{d^2 a^\prime}{\pi} \, \mathrm{e}^{\mathrm{Re}(a a^\prime)} K_0(|a a^\prime|)
    \, \Delta Y_{\rm sc} (a^\prime) \right)^{K}
\end{align}
where $w^2 = w_V^2 / w_T^2$ and the variables have been scaled as $a \to a / \sqrt{2} w_T$. The modified Bessel function $K_0$ is due to integration over $\varphi^\prime$, and the AAT equation is closed by taking $\Delta = - \partial^2 / \partial a \partial \bar{a}$. An important remark here is that $x_1 = \mathrm{Re}(a)$ parameterizes an $\mathrm{SO}(1,2)$-symmetry direction, as it arises from the expression (\ref{eq:Q-cart}) by starting with $x_1 = 0$, $x_2 = (x_0^2 - 1)/2$ and then conjugating $Q \mapsto g Q g^{-1}$ with $g = \mathrm{e}^{ \mathrm{Re}(a) (\mathrm{i}\sigma_3 - \sigma_2)/2} \in \mathrm{SU}(1,1)$ to generate a horocycle. The term $(w_\beta^2 / w_T^2) \, \mathrm{Re}^2 (a)$ in the exponent of Eq.\ (\ref{eq:eqn4Ysc}) explicitly breaks that horocycle symmetry.

\begin{figure}
    \centering
    \includegraphics[width=7cm]{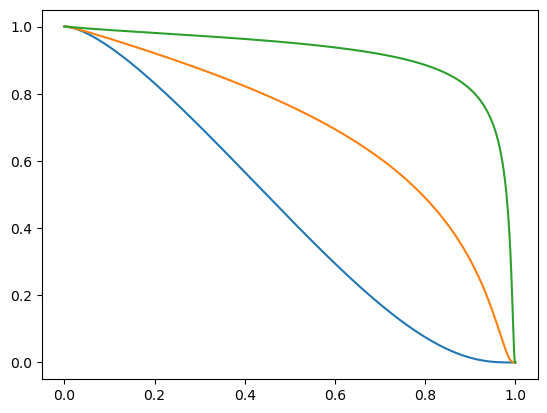}
    \caption{Graphs of the solution $Y_{\rm sc}(a)$ along $\mathrm{arg}(a) = 0$ as a function of  $r = |a|/(1+|a|)$, for tree branching number $K = 4$ and varying values of the disorder parameter $w^2$. We find that $Y_{\rm sc}$ is ac-unstable for $w^2 = 2$ and $w^2 = 20$ (blue and orange graphs) but stable for $w^2 = 100$ (green graph).} \label{fig:F2}
\end{figure}

We have solved Eq.\ (\ref{eq:eqn4Ysc}) numerically by discretizing $|a^\prime|$ and expanding in Fourier modes $\mathrm{e}^{\mathrm{i} k\, \mathrm{arg}(a^\prime)}$ ($k \in \mathbb{Z}$). The graph of the solution $Y_{\rm sc}$ for the case of $K = 4$ and no explicit symmetry breaking ($w_\beta^2 = 0$) is plotted in Fig.\ \ref{fig:F2} for  $\mathrm{arg}(a) = 0$ and various values of $w^2$. We find that $Y_{\rm sc}$ exists as a stable solution with spontaneously broken symmetry in an interval $w_{c1} < w < w_{c2}\,$. At the upper end $(w = w_{c2})$, $Y_{\rm sc}$ turns into the pp-type solution, which is independent of the $G$-symmetry variable $\mathrm{Re}(a)$. At the lower end $(w = w_{1c})$, $Y_{\rm sc}$ becomes unstable with respect to fluctuations of the Lorentz boost variable $\varphi$, marking the onset of full symmetry breaking in the ac-phase.

To throw further light on $Y_{\rm sc}$, we note the expression
\begin{equation}\label{eq:avGsqu}
    \mathbb{E} (|G_n|^2) = \int d\mu(Q) \left( x_0 \frac{\partial}{\partial x_0} + x_3 \frac{\partial}{\partial x_3} \right) Y(Q)
\end{equation}
for the disorder average of the squared local Green's function. (Here we take $x_2 = \pm \sqrt{x_3^2 - x_0^2 - x_1^2}$ with coordinates
$x_3, x_0, x_1$.) Evaluation of the integral (\ref{eq:avGsqu}) gives a finite result for $Y = Y_{\rm ac}$ and a pole singularity ($\sim \varepsilon^{-1}$) for $Y = Y_{\rm pp}$. In the case of $Y = Y_{\rm sc}$ we obtain a singularity
\begin{equation}
    \mathbb{E} (|G_n|^2) \sim \varepsilon^{-\alpha} \quad (0 < \alpha < 1) ,
\end{equation}
weakened in degree by just one non-compact symmetry (instead of two) being unbroken. Such behavior is indicative of a singular continuous spectral measure \cite{AK2023}. 

In summary, the solution $Y_{\rm sc}$ preserves a Lorentz boost symmetry but spontaneously breaks a horocycle symmetry (and hence the symmetry under $\mathrm{SO}(2)$ rotations in the $x_1 x_2$ plane) of $G^\prime = \mathrm{SO}(1,2)$. Thus, our numerical results clearly demonstrate the existence, for the $N=1$ Wegner model, of a non-standard phase (sc) distinct from the two established phases (pp and ac). To explain the paper's title, we note that the compact symmetry $\mathrm{U}(1) \subset G$, cf.\ Eq.\ (\ref{eq:U(1)}), acts on $Q$ as $\mathrm{SO}(2) \subset G^\prime$.

\section{Conclusion}

By solving the $N=1$ Wegner model in an approximation exact on the Bethe lattice and likely valid in high dimension, we have demonstrated the existence of a non-standard phase (sc) of disordered electronic matter. The new phase (sc) is set apart from the conventional phases (pp and ac) by three characteristic features: (i) a spectral measure of singular continuous type, (ii) energy eigenstates with fractal support, and (iii) a distinct form of partial symmetry breaking in the field-theory formalism. The latter scenario is non-standard in that a compact symmetry such as the $\mathrm{U}(1)$ of Eq.\ (\ref{eq:U(1)}), is spontaneously broken while a non-compact symmetry of Lorentzian type remains unbroken. First discovered in the quest of identifying the conformal field theory of the integer quantum Hall transition \cite{CFT-IQHT}, that scenario is likely to rule the scaling behavior of a broad variety of Anderson transitions at strong coupling.

In order to arrive with relative ease at the flow diagram conjectured in Fig.\ \ref{fig:F1}, one may want to replace the AAT scheme by an approximate renormalization group (RG) scheme, e.g.\ the Dyson hierarchical scheme or the Migdal-Kadanoff approximation. Combining our analytical insight with  numerical simulations \cite{Lemarie2022, SLS2022}, we expect the RG flow to be attracted to a field theory with two relevant parameters, one ($\lambda_\tau$) controlling the field fluctuations tangential to the Lorentz boost directions of the light cone ($x_3^2 - x_1^2 - x_2^2 = 0$), the other ($\lambda_\sigma$) controlling the fluctuations of the $x_1 x_2$ rotational field degrees of freedom (or rather, supersymmetry analogs thereof).

\smallskip\noindent{\bf Acknowledgment.} JA acknowledges financial support by the DFG-Sonderforschungsbereich CRC 183 and the Center QM2 of the University of Cologne.

\end{document}